\documentclass{cernphprep}

\usepackage{graphicx}
\DeclareGraphicsExtensions{.eps,.pdf,.png,.gif,.jpg,.mht}
\usepackage{multirow}
\usepackage{amsfonts}
\usepackage{amstext}
\usepackage{amsmath}
\usepackage{afterpage}

\usepackage{amssymb}
\usepackage{verbatim}

\begin{document}
\renewcommand{\thefootnote}{\fnsymbol{\footnote}}
\begin{titlepage}
\PHnumber{2011-158}
\PHdate{24 September 2011}
\EXPnumber{TOTEM 2011--02}


\title{First measurement of the total proton-proton cross section at the LHC energy of $\mathbf{\sqrt s =\,}$7\,TeV}

\Collaboration{The TOTEM Collaboration 
}
\ShortAuthor{The TOTEM Collaboration}
\ShortTitle{Proton-proton total cross section at LHC}



\begin{Authlist}
G.~Antchev\footnote{INRNE-BAS, Institute for Nuclear Research and Nuclear Energy, Bulgarian Academy of Sciences, Sofia, Bulgaria}\addtocounter{footnote}{-1}, P.~Aspell$^{8}$, I.~Atanassov$^{8,}$\footnotemark, V.~Avati$^{8}$, J.~Baechler$^{8}$,
V.~Berardi$^{5b,5a}$, M.~Berretti$^{7b}$, E.~Bossini$^{7b}$, M.~Bozzo$^{6b,6a}$, P. Brogi$^{7b}$ ,
E.~Br\"{u}cken$^{3a,3b}$, A.~Buzzo$^{6a}$, F.~Cafagna$^{5a}$, M.~Calicchio$^{5b,5a}$,
M.~G.~Catanesi$^{5a}$, C.~Covault$^{9}$,
T.~Cs\"{o}rg\H{o}$^{4}$,
M.~Deile$^{8}$,
K.~Eggert$^{9}$, V.Eremin\footnote{Ioffe Physical - Technical Institute of Russian Academy of Sciences}, R. Ferretti$^{6a, 6b}$, F.~Ferro$^{6a}$, A. Fiergolski\footnote{Warsaw University of Technology, Poland}, F.~Garcia$^{3a}$, S.~Giani$^{8}$,
V.~Greco$^{7b,8}$, L.~Grzanka$^{8,}$\footnote{Institute of Nuclear Physics, Polish Academy of Sciences, Cracow, Poland}\addtocounter{footnote}{-2}, J.~Heino$^{3a}$, T.~Hilden$^{3a,3b}$, M.R. Intonti$^{5a}$,
J.~Ka\v{s}par$^{1a,8}$, J.~Kopal$^{1a,8}$, V.~Kundr\'{a}t$^{1a}$, K.~Kurvinen$^{3a}$,
S.~Lami$^{7a}$, G.~Latino$^{7b}$,
R.~Lauhakangas$^{3a}$, T. Leszko\footnotemark\addtocounter{footnote}{+1},
E.~Lippmaa$^{2}$,
M.~Lokaj\'{\i}\v{c}ek$^{1a}$, M.~Lo~Vetere$^{6b,6a}$, F.~Lucas~Rodr\'{\i}guez$^{8}$,
M.~Macr\'{\i}$^{6a}$, L.~Magaletti$^{5b,5a}$,
A.~Mercadante$^{5b,5a}$,
S.~Minutoli$^{6a}$, F.~Nemes$^{4,}$\footnote{Department of Atomic Physics, ELTE University, Budapest, Hungary},
H.~Niewiadomski$^{8}$,
E.~Oliveri$^{7b}$, F.~Oljemark$^{3a,3b}$, R.~Orava$^{3a,3b}$, M.~Oriunno$^{8}$\footnote{SLAC National Accelerator Laboratory, Stanford CA, USA},
K.~\"{O}sterberg$^{3a,3b}$,
P.~Palazzi$^{7b}$,
J.~Proch\'{a}zka$^{1a}$, M.~Quinto$^{5a}$,
E.~Radermacher$^{8}$, E.~Radicioni$^{5a}$,
F.~Ravotti$^{8}$, E.~Robutti$^{6a}$,
L.~Ropelewski$^{8}$, G.~Ruggiero$^{8}$,
H.~Saarikko$^{3a,3b}$, G.~Sanguinetti$^{7a}$,  A.~Santroni$^{6b,6a}$,
A.~Scribano$^{7b}$,
W.~Snoeys$^{8}$,
J.~Sziklai$^{4}$, C.~Taylor$^{9}$,
N.~Turini$^{7b}$, V.~Vacek$^{1b}$, M.~V\'{i}tek$^{1b}$, J.~Welti$^{3a,b}$, J.~Whitmore$^{10}$.\\
\vspace{0.5cm}
$^{1a}${Institute of Physics, Academy of Sciences of the Czech Republic, Praha, Czech Republic.}\\
$^{1b}${Czech Technical University, Praha, Czech Republic.}\\
$^{2}${National Institute of Chemical Physics and Biophysics NICPB, Tallinn, Estonia.}\\
$^{3a}${Helsinki Institute of Physics, Finland.}\\
$^{3b}${Department of Physics, University of Helsinki, Finland.}\\
$^{4}${MTA KFKI RMKI, Budapest, Hungary.}\\
$^{5a}${INFN Sezione di Bari, Italy.}\\
$^{5b}${Dipartimento Interateneo di Fisica di Bari, Italy.}\\
$^{6a}${Sezione INFN, Genova, Italy.}\\
$^{6b}${Universit\`{a} degli Studi di Genova, Italy.}\\
$^{7a}${INFN Sezione di Pisa, Italy.}\\
$^{7b}${Universit\`{a} degli Studi di Siena and Gruppo Collegato INFN di Siena, Italy.}\\
$^{8}${CERN, Geneva, Switzerland.}\\
$^{9}${Case Western Reserve University, Dept. of Physics, Cleveland, OH, USA.}\\
$^{10}${Penn State University, Dept. of Physics, University Park, PA, USA.}\\

\end{Authlist}
\vspace{2cm}

\begin{abstract}
TOTEM has measured the differential cross-section for elastic proton-proton
scattering at the LHC energy of $\sqrt{s} = 7\,$TeV analysing data from a short
run with dedicated large $\beta^{*}$ optics. A single exponential fit with
a slope $B = (20.1 \pm 0.2^{\rm stat} \pm 0.3^{\rm syst})\,\rm GeV^{-2}$ describes the
range of the four-momentum transfer squared $|t|$ from 0.02 to 0.33\,GeV$^{2}$.
After the extrapolation to $|t|=0$, a total elastic scattering cross-section
of $(24.8 \pm 0.2^{\rm stat} \pm 1.2^{\rm syst})$\,mb was obtained.
Applying the optical theorem and using the
luminosity measurement from CMS, a total proton-proton cross-section of
$(98.3 \pm 0.2^{\rm stat} \pm 2.8^{\rm syst})$\,mb was deduced which is in good
agreement with the expectation from the
overall fit of previously measured data over a large range of center-of-mass energies. From
the total and elastic pp cross-section measurements, an inelastic
pp cross-section of $(73.5 \pm 0.6^{\rm stat} ~^{+1.8}_{-1.3}~^{{\rm syst}})$\,mb was
inferred.
\end{abstract}
\vspace{2cm}
\begin{center}
    {\em Accepted for publication in EPL}
\end{center}

\end{titlepage}







\section{Introduction}
The observation of the rise of the total cross-section with energy was one of the highlights at the ISR, the  first CERN collider~\cite{Amaldi:1973yv,Amendolia:1973yw,Baksay:1978sg,Amaldi:1978vc}.
Some indirect indications for this unforeseen phenomenon had already come earlier from high-energy cosmic ray showers~\cite{Akimov:1969zzb,Yodh:1972fv,Dawson:1987iv}.
A long series of total proton-antiproton cross-section measurements followed in the last decades both at the CERN Sp$\bar{\rm p}$S collider~\cite{Battiston:1982su,Arnison:1983mm} and at the TEVATRON~\cite{Amos:1989at,Abe:1993xy,Sadr:1993jq,Avila:1998ej}.

In this letter, we report the first measurement of the total and elastic proton-proton cross-sections at the CERN Large Hadron Collider (LHC) using the optical theorem together with the luminosity provided by the CMS experiment~\cite{int:lumi1,int:lumi2}.
With a dedicated beam optics configuration ($\beta^{*} = 90$\,m) TOTEM has measured the differential cross-section of elastic scattering for four-momentum transfer squared values
$|t|$ to $2\times 10^{-2}\,\rm GeV^{2}$, making the extrapolation to the optical point at $|t| = 0$ possible.
This allows the determination of the elastic scattering cross-section as well as the total cross-section.

\section{The Roman Pot detectors}
To detect leading protons scattered at very small angles, silicon sensors are placed in movable beam-pipe insertions -- so-called ``Roman Pots'' (RP) -- located symmetrically on either side of the LHC interaction point IP5 at distances of 215 -- 220\,m from the IP~\cite{Anelli:2008zza}.

Each RP station is composed of two units separated by a distance of about 5\,m.
A unit consists of 3 RPs, two approaching the outgoing beam vertically and one horizontally.
Each RP is equipped with a stack of 10 silicon strip detectors designed with the specific objective of reducing the insensitive area at the edge facing the beam to only a few tens of micrometers.
The 512 strips with $66\,\mu$m pitch of each detector are oriented at an angle of $+45^{\circ}$ (five ``$u$''-planes) and $-45^{\circ}$ (five ``$v$''-planes) with respect to the detector edge facing the beam.
During the measurement the detectors in the horizontal RPs overlap with the ones in the vertical RPs, enabling a precise ($10\,\mu$m) relative alignment of all three RPs in a unit by correlating their positions via common particle tracks.
The precision and the reproducibility of the alignment of all RP detector planes with respect to each other and to the position of the beam centre is one of the most delicate and difficult tasks of the experiment~\cite{IPAC:2011}.

In a station, the long lever arm between the near and the far RP units has two
important advantages.
First, the local track angles in the x- and y-projections perpendicular to the
beam direction are reconstructed with a precision of 5 to $10\,\mu$rad.
Second, the proton trigger selection by track angle can use all RPs
independently: the stations on the opposite sides of the IP,
the near and far units of each station, and the $u$ and $v$ planes in each unit.
This redundancy can be used to obtain high trigger efficiency or purity.

\section{The special LHC optics}
After an elastic scattering interaction in IP5 with the transverse vertex position $(x^{*}, y^{*})$ and with scattering angle projections $(\Theta_{x}^{*}, \Theta_{y}^{*})$,
the displacement $(x, y)$ of the proton trajectory from the beam centre at the RP position $s_{RP}$ is given by:
\begin{equation}
x = L_{x} \Theta^{*}_{x} + v_{x} x^{*} \:, \quad
y = L_{y} \Theta^{*}_{y} + v_{y} y^{*} \:,
\end{equation}
where the optical functions $L_{x, y}$ and $v_{x, y}$ at the RP position $s_{\rm RP}$ are determined by the beta function:
$L_{x, y} = \sqrt{\beta_{x, y} \beta^*}\, \sin\Delta\mu_{x, y}$ and $v_{x, y} = \sqrt{\frac{\beta_{x, y}}{\beta^*}}\, \cos\Delta\mu_{x, y}$ with the phase advance $\Delta \mu_{x, y} = \int_{\rm IP}^{s_{\rm RP}} \frac{1}{\beta_{x, y}(s)}\,{\rm d}s$ relative to the IP.
To maximise the sensitivity of the position measurement to the scattering angle while minimising its dependence on the vertex position, special optics are designed to have minimum beam divergence $\sigma_{\Theta}^{*}$ at the IP (imposing large values of $\beta^{*}$ via $\sigma_{\Theta}^{*} = \sqrt{\varepsilon_{n}/\beta^*}$), large values of $L$ and $v = 0$, and thus $\Delta \mu = \pi/2$ in at least one projection.
In the ultimate TOTEM optics with $\beta^{*} = 1540\,$m~\cite{Verdier:2005av} this so-called ``parallel-to-point focussing'' condition will be fulfilled in both $x$ and $y$.
As a first step towards this goal, the intermediate optics with $\beta^{*} = 90\,$m was developed~\cite{EggertBlois:2005,Optics:90memo,Burkhardt:2010sw}.
Since this intermediate optics is reachable by gradually increasing $\beta^{*}$ from 11\,m (the value at injection) to 90\,m (``un-squeezing''), the commissioning is easier than for the ultimate optics.
The $\beta^{*} = 90\,$m exhibits parallel-to-point focussing only in the vertical plane ($\Delta \mu_{y} \approx \pi/2$, $L_{y} \approx 260\,$m, $v_y \approx 0$), whereas in the horizontal plane $\Delta \mu_{y} \approx \pi$ and hence $L_x \approx 0$, which helps separating elastic and diffractive events.
The beam divergence is $\sigma_{\Theta}^{*} \approx 2.5 \mu$rad.
The vertical scattering angle $\Theta_{y}^{*}$ can then be directly reconstructed from the track position $y$, whereas due to $L_{x} \approx 0$ the horizontal component $\Theta_{x}^{*}$ is optimally reconstructed from the track angle $\Theta_{x} = \frac{{\rm d}x}{{\rm d}s}$ at the RP:
\begin{equation}
\Theta^{*}_{y} = \frac{y}{L_{y}} \:, \quad
\Theta^{*}_{x} = \frac{1}{\frac {{\rm d}L_x}{{\rm d}s}} \: \left( \Theta_x - {\frac{{\rm d}v_x}{{\rm d}s}}\, x^{*} \right) \: ,
\end{equation}
where the unmeasured vertex produces a smearing term $\propto x^{*}$ .  However this smearing is eliminated later in the analysis since the vertex term cancels due to the correlation between the collinear tracks of the two outgoing protons.

\section{Data collection and event selection}
\begin{figure}[ht!]
\begin{center}
\includegraphics[width=0.45\textwidth]{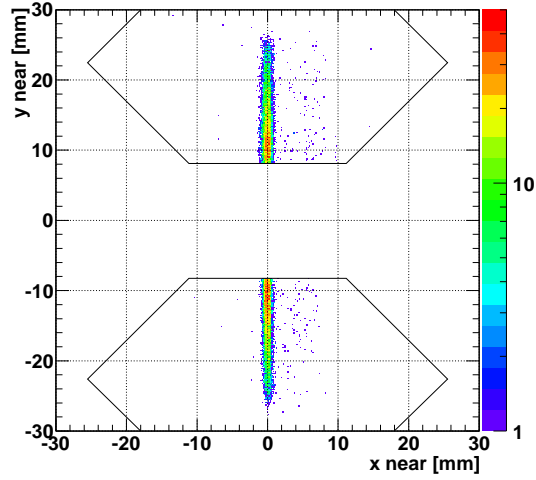}
\end{center}
\caption{The intersection points of all reconstructed tracks in this data set with the RP silicon detectors (black lines indicate the detector sensitive area) at the RP\,220\,m station. In order to represent the pp scattering configuration the tracks visualized in the bottom silicon detector refer to one side and in the top detector to the other side of the interaction point.}
\label{fig_reco-tracks}
\end{figure}

The data presented here were collected in the first LHC run with the $\beta^{*} = 90\,$m optics. Each beam had two bunches with populations of $1 \times 10^{10}$ and $2 \times 10^{10}$ protons.
Given the normalised transverse emittances of $(1.8\div2.6)\,\mu$m\,rad depending on the bunch, this filling scheme led to an instantaneous luminosity of about $8 \times 10^{26}\,\rm cm^{-2} s^{-1}$.
Thanks to the low beam intensity, the RP detectors could safely approach the beam centre to a distance of 10 times the transverse beam size.
After verifying that the beam orbit did not significantly differ from the one with nominal beam optics ($\beta^{*} = 1.5\,$m), the RP positions were defined relative to the reference beam centre determined one month earlier in a beam-based alignment exercise for $\beta^{*} = 1.5\,$m.
Within the running time of 33 minutes, an integrated luminosity of
$1.7\,\mu\rm b^{-1}$ was delivered, and 66950 events
were recorded with a very loose trigger requiring a track segment in any of the vertical RPs in at least one of the two transverse projections ($u$, $v$).
The data sample relevant for this analysis consisted of 15973 events characterised by the elastic double-arm signature in the vertical RPs ({\it top left of IP - bottom right of IP} or {\it bottom left of IP - top right of IP}).
Fig.~\ref{fig_reco-tracks} shows the intersection points of the selected tracks with the RP detectors at the RP\,220\,m stations on both sides of the IP.

%
\begin{figure}[h!]
\begin{center}
\includegraphics[width=0.49\textwidth]{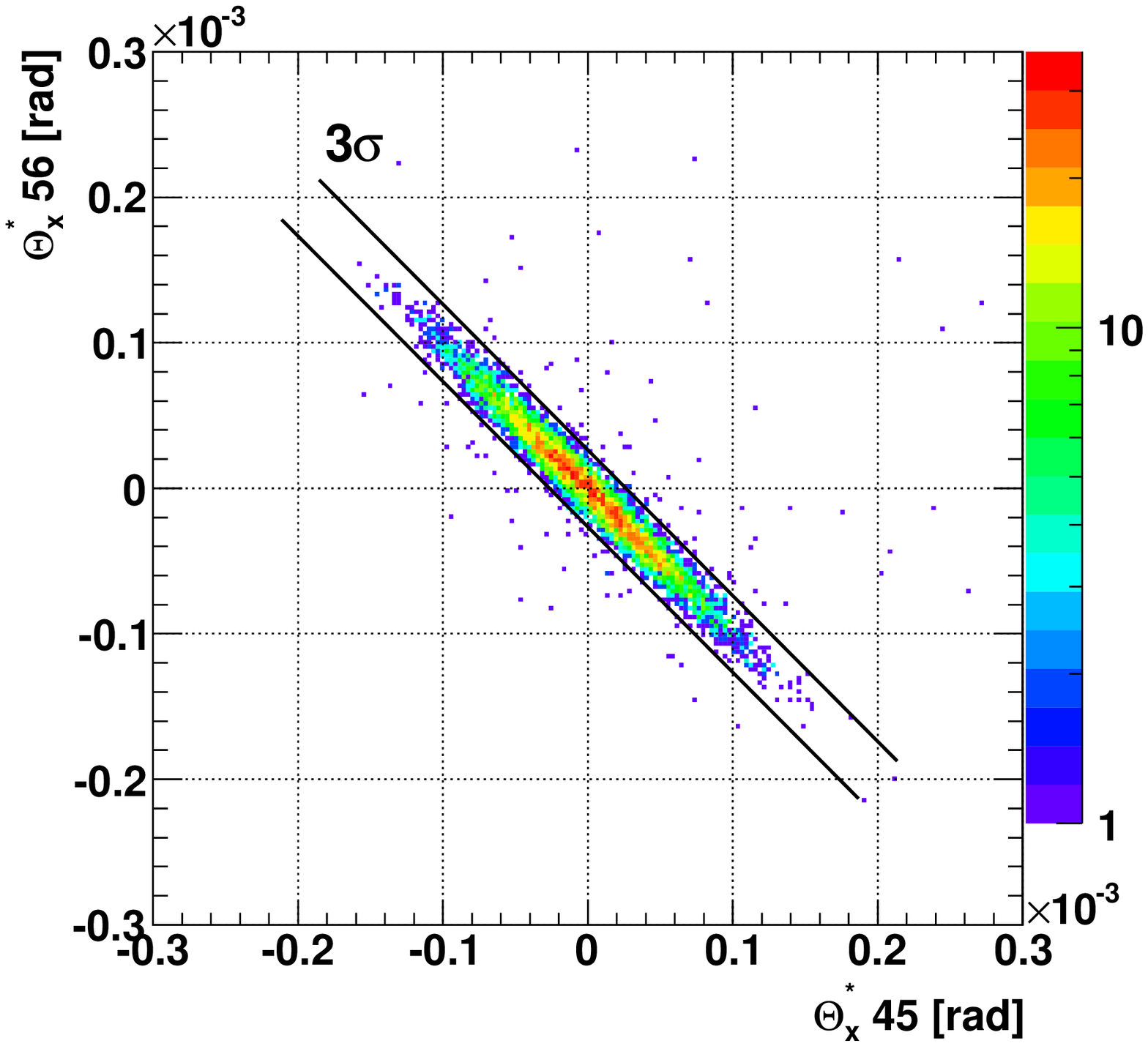}
\hfill
\includegraphics[width=0.49\textwidth]{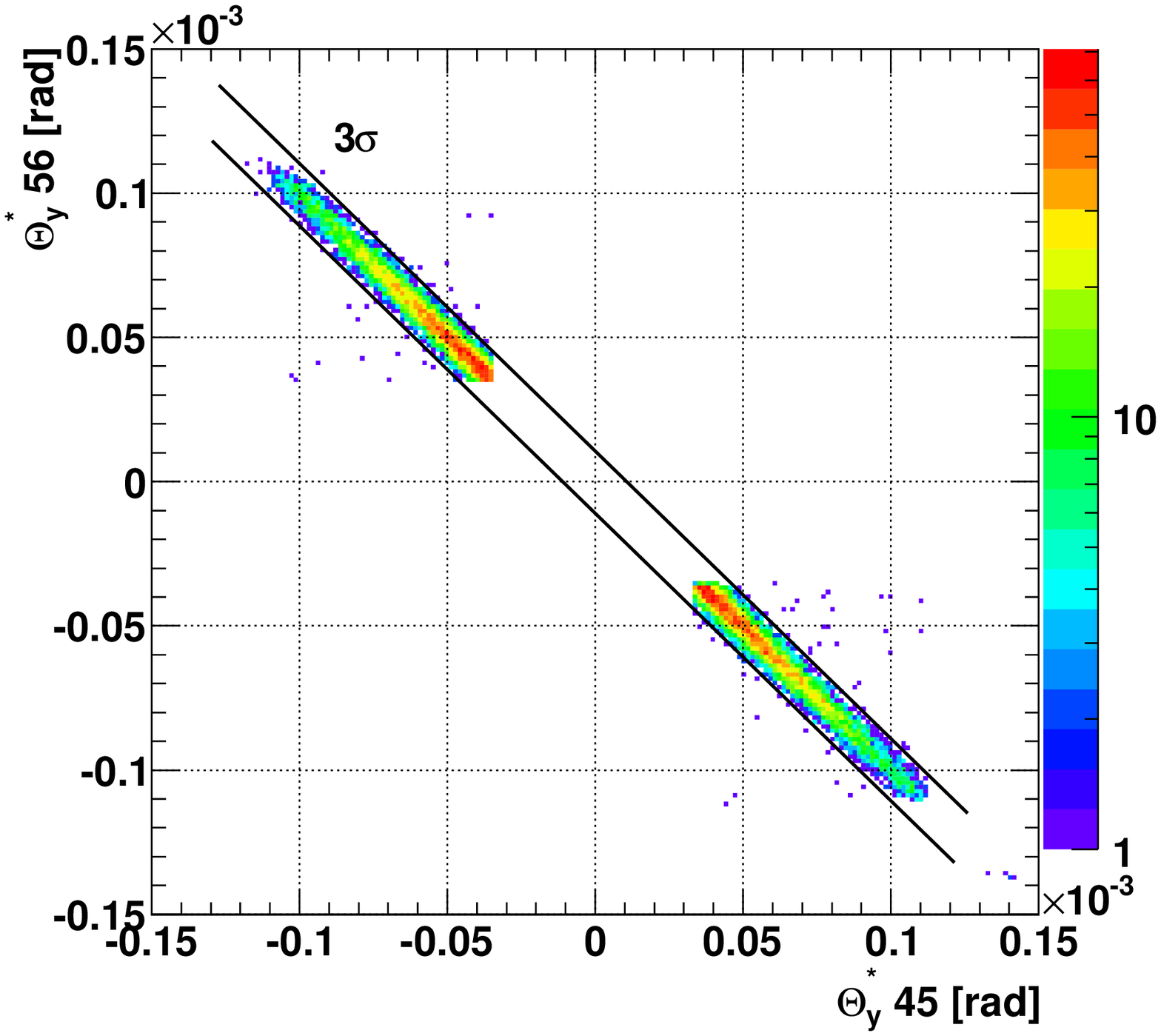}
\end{center}
\caption{The correlation between the reconstructed proton scattering angles $\Theta^{*}_{x}$ (left plot) and $\Theta^{*}_{y}$ (right plot) on both sides of the IP (``45" = left of IP5, ``56" = right of IP5).}
\label{fig_angle-correl}
\end{figure}

%
\section{Analysis}
\paragraph{Elastic tagging}
Due to the purity of the data obtained in the conditions of this special run, the final refinement of the elastic event selection -- requiring collinearity of the two outgoing protons reconstructed with full detector efficiency within 3 standard deviations in their scattering angle correlation (Fig.~\ref{fig_angle-correl}) -- reduces the sample to 14685 elastic events.
No further cuts, e.g. for excluding diffractive events, are necessary.
The reconstructed and selected elastic events from the two allowed diagonal topologies (7315 events on {\it top left of IP - bottom right of IP} and 7370 events on
{\it bottom left of IP - top right of IP}) showed that their acceptance was the same and that the RP system was well aligned.
The $\Theta^{*}_{y}$ resolution was $1.7\,\mu$rad, originating directly from the beam divergence since detector effects are suppressed given the large value of $L_y$.
In the $\Theta^{*}_{x}$ resolution the beam divergence contribution is convoluted with the detector resolution and the vertex distribution, but the vertex effect is factorised out once the pairs of elastic protons are reconstructed together.

\paragraph{Optics, t-scale}
The LHC optics with $\beta^* = 90$\,m is very insensitive to machine parameter
variations. This led to systematic uncertainties on $\Theta^{*}_{x}$
and $\Theta^{*}_{y}$ of just 1.3\% and 0.4\%, respectively.
Non-linearities were observed and corrected in the observable $\Theta^{*}_{x}$ as a function of the reconstructed $y$ position.
The correction was benchmarked on the isotropy of the $\phi$ distribution of the elastic candidate events, and it was cross-checked by repeating the analysis using $L_x$ to reconstruct the relevant physics variables.
The overall propagated systematic uncertainty on the $t$ scale for one arm
is 0.8\% at low $|t|$ and 2.6\% at large $|t|$.

\begin{figure*}[h!]
\begin{center}
\includegraphics[width=0.50\textwidth]{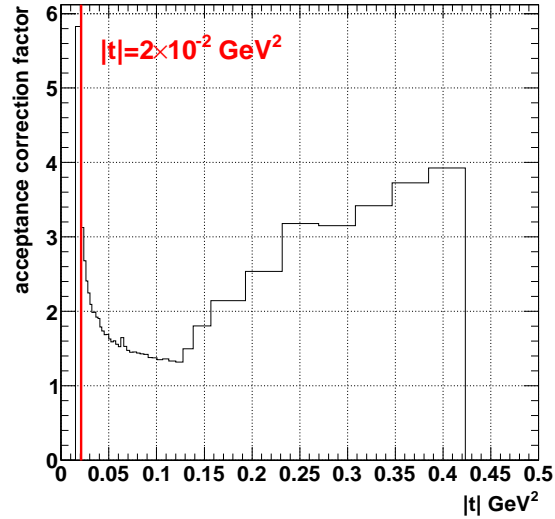}
\end{center}
\caption{The acceptance correction factor and the $t$-limit of this analysis.}
\label{fig_acceptance-corrections}
\end{figure*}

\paragraph{Acceptance}
The acceptance limitations at low $|t|$ have been corrected for the geometry and beam divergence related factors (Fig.~\ref{fig_acceptance-corrections}).
%
The TOTEM experiment has thus measured $|t|$ down to the limit of $2 \times 10^{-2}\,\rm GeV^2$ (constraining the total acceptance correction at low $|t|$ to be $\leq 3$ in order to minimise the systematics).
The acceptance loss (due to aperture limitations) at the high end of the
$|t|$ spectrum measurable with this optics has also been treated, thus
allowing measurements of $|t|$ up to $0.42\,\rm GeV^2$.
Therefore a comparative analysis with the previously
published TOTEM results~\cite{EPLelastic:2011} is possible.

\paragraph{Efficiency}
The detector and reconstruction efficiency per pot has been evaluated directly from the data, by repeating the analysis of the elastic events using three pots out of four and permutating on the missing pot. The inefficiency of the near pots was about 1.5\%, while the inefficiency of the far pots was about 3\%\,: such a difference is expected due to the probability of a proton having a hadronic interaction in a near pot, inducing a shower onto its corresponding far pot.
The overall inefficiency for both diagonals has been computed to be 8.9\% and 8.7\%, counting also the uncorrelated probabilities to have more than one pot inefficient at the same time.
The special trigger combination with all detector components in OR, used for this data taking, has allowed checking and excluding all combinations of correlated inefficiencies.

\paragraph{Background}
The data did not show any measurable background ($< 0.1$\%) affecting the selection of elastic events.
Single Gaussian fits precisely describe (without any non-Gaussian tails) the distributions resulting from the selection cuts, guaranteeing efficiency and purity of these cuts.
In fact, the special optics run and the two colliding low-intensity bunches
ensured the absence of pile-up from single diffraction; moreover, the data
have shown that double pomeron exchange events
could not satisfy the
collinearity requirements in both dimensions at the same time even at very
low $|t|$, as verified by selecting events with momentum loss $\Delta p/p > 1$\%.

\paragraph{Resolution}
After deconvolution of the vertex, the effective resolution in $\Theta^{*}_{x}$  has contributions of $1.7\,\mu$rad from the beam divergence and $4.0\,\mu$rad from the RP detector resolution.
The resolution in $\Theta^{*}_{y}$ is $1.7\,\mu$rad from the beam divergence.
Hence the bin migration correction as function of $t$ was contained between $+1$\% and $-3$\%.
This resolution-unfolding correction has been computed with high precision (systematic uncertainty of 0.7\%) taking into account the acceptance effects, given the different resolutions in the two angular components.

\paragraph{Extrapolation to {\em  t = 0}}
The elastic differential cross-section has been measured down to
$|t| = 2 \times 10^{-2}\,\rm GeV^{2}$. The data were then extrapolated to $t = 0$
assuming the functional form
\begin{equation}
\frac{{\rm d}\sigma_{\rm el}}{{\rm d}t} = \left. \frac{{\rm d}\sigma_{\rm el}}{{\rm d}t}\right|_{t=0} {\rm e}^{-B\,|t|} \: .
\end{equation}
The statistical and the propagated systematic uncertainties of the extrapolation
are given separately in Tab.~\ref{tab_results}.

\paragraph{Luminosity, Trigger, Normalisation}
This data analysis was based on a large fraction of the data taken during
run 5657, excluding the period of initial beam adjustment, in fill 1902 on
29 June 2011. The luminosity was recorded by CMS with an uncertainty of
4\%~\cite{int:lumi1,int:lumi2}; the
additional uncertainty contribution due to pile-up, found by CMS in
2011~\cite{cmslumi:2011} does not apply to the very low luminosity discussed
here.
The trigger efficiency for elastic events
was greater than 99.9\%.
Thus the effective integrated luminosity was equal to $1.65\,\mu\rm b^{-1}$.

\section{Results}
\begin{figure*}[!hb]
\begin{center}
\includegraphics[width=0.90\textwidth]{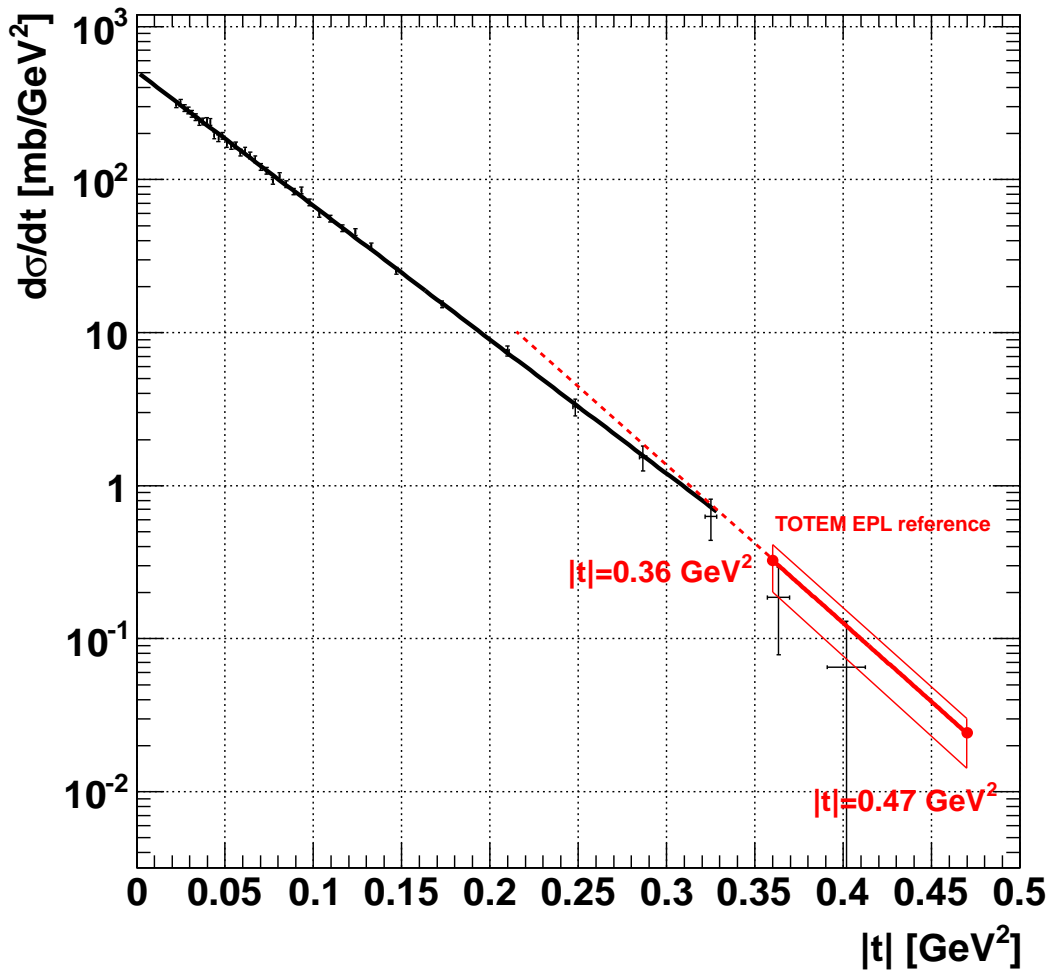}
\end{center}
\caption{The measured pp elastic scattering differential cross-section
$\rm{d}\sigma/\rm{d}t$. The superimposed fits and their parameter values are
discussed in the text.}
\label{fig_final-t-distribution}
\end{figure*}

After including all the analysis corrections described above, the final
differential cross-section for the elastic proton-proton scattering with its
statistical errors is shown in Fig.~\ref{fig_final-t-distribution}.

The new data can be described by a single exponential fit ($\chi^{2}/\rm d.o.f. = 0.8$) over the complete $|t|$ range of $(0.02 \div 0.33)\,\rm GeV^2$ with the slope $B = (\rm 20.1 \pm 0.2 (stat) \pm 0.3 (syst))\,GeV^{-2}$.
The TOTEM result for $B$ at low $|t|$ confirms the trend of an increase with
$\sqrt{s}$~\cite{Amos:1989at,Guryn:2004,Abe:1993xx,Amos:1988ng,Amos:1990fw,Amos:1992zn,Brandt:2010zz}.
As a comparison, the exponential fit at the lower end of the $|t|$ range of our previous measurement~\cite{EPLelastic:2011} which covered $(0.36 \div 2.5)\,\rm GeV^2$ is also shown in Fig~\ref{fig_final-t-distribution}. The agreement between the two measurements that were done with different optics is excellent.
It is worth noting that the slope in the $|t|$ interval of $(0.36 \div 0.47)\,\rm GeV^{2}$ is significantly larger: $(\rm 23.6 \pm 0.5 (stat) \pm 0.4 (syst))\,GeV^{-2}$.

Assuming a constant slope $B$ for the nuclear scattering, the differential cross-section at the optical point $t=0$ was determined to be $\frac{{\rm d}\sigma}{{\rm d}t}|_{t=0} =\rm (503.7 \pm 1.5 (stat) \pm 26.7 (syst))\,mb / GeV^{2}$. Integrating the differential cross-section yields a total elastic scattering cross-section of $\rm (24.8 \pm 0.2 (stat) \pm 1.2 (syst))\,mb$, out of which 16.5\,mb were directly observed.

The total proton-proton cross-section is related to the elastic cross-section via the optical theorem
\begin{equation}
\sigma_{\rm tot}^{2} = \frac{16 \pi (\hbar c)^{2}}{1 + \rho^{2}}
\left.\frac{{\rm d}\sigma_{\rm el}}{{\rm d}t} \right|_{t=0} \: .
\end{equation}
Taking the COMPETE prediction~\cite{Cudell:2002} of $0.14^{+0.01}_{-0.08}$ for
the parameter $\rho = \frac{\mathcal{R}[f_{\rm el}(0)]}{\mathcal{I}[f_{\rm el}(0)]}$,
where $f_{\rm el}(0)$ is the forward nuclear elastic amplitude,
$\sigma_{\rm tot}$ was thus determined to be
\begin{equation}
\sigma_{\rm tot} = \left(\rm 98.3 \pm 0.2 (stat) \begin{array}{c}+2.8\\-2.7\end{array} (syst)\right)\,\rm mb.
\end{equation}
The errors are dominated by the extrapolation to $t=0$ and the luminosity uncertainty.

Subtracting the elastic scattering cross-section we obtain a value for the inelastic cross-section which can then be compared with the measurements of the CMS~\cite{CMSint:2011}, ATLAS~\cite{Aad:2011eu}, and ALICE~\cite{ALICE:2011}
experiments.
The results (Tab.~\ref{tab_results}) are consistent within the quoted errors
of CMS, ATLAS, and ALICE, which took into account the uncertainties of the
model predictions for the unobserved very-forward diffractive processes.

\begin{figure*}[!htb]
\begin{center}
\includegraphics[width=\textwidth]{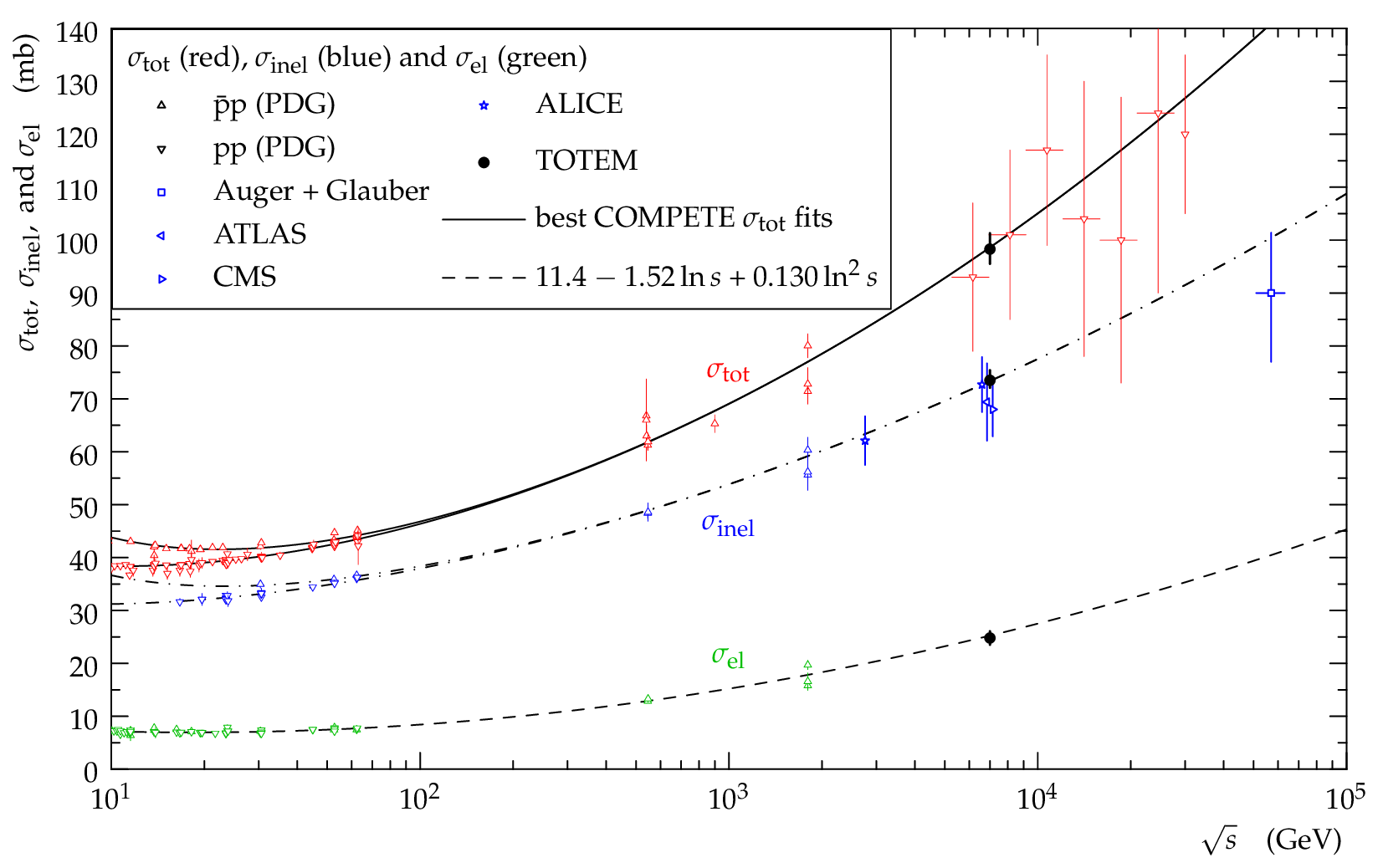}

\end{center}

\caption{Compilation of total ($\sigma_{\rm tot}$), inelastic ($\sigma_{\rm inel}$) and elastic ($\sigma_{\rm el}$) cross-section measurements~\cite{Cudell:2002,CMSint:2011,Aad:2011eu,ALICE:2011,PDG:2010,Block:2011nr}.}
\label{fig_compilation}
\end{figure*}

In Fig.~\ref{fig_compilation}, the values of the TOTEM total and elastic
cross-sections are compared with results at lower energies and from cosmic
rays together with an overall fit of the COMPETE collaboration~\cite{Cudell:2002}. The TOTEM total cross-section is in excellent agreement with the extrapolation from lower energies. To guide the eye, a parabolic fit was used for the energy dependence of the elastic cross-section. The ratio of the elastic to total cross-section,
$\sigma_{\rm el}/\sigma_{\rm tot} = 0.25 \pm 0.01^{\rm stat \oplus syst}$.

Table~\ref{tab_results} lists the values of the measured observables
and the final results for the physics quantities along with
their statistical and systematic uncertainties.

\begin{table*}[!ht]
\begin{center}
\caption{Results of the TOTEM measurements at the LHC energy of $\sqrt{s}= 7\,$TeV.}
\renewcommand{\arraystretch}{1.3}  
\scalebox{0.8}{
\begin{tabular}{|l|l|l|l|}
  \hline
  & Statistical uncertainties & Systematic uncertainties & Result \\
  \hline
  $t$ & $\pm [3.4\div11.9]\% $
    & $\pm[0.6\div1.8]\% ^{\rm optics} \pm <1 \%^{\rm alignment}$ & \vspace{-2.5truemm}\\
    & \footnotesize{single measurement}$^{(*)}$ &  & \\
   \hline
$\frac{{\rm d}\sigma}{{\rm d}t}$  &  5\% / bin
    &  $\pm 4 \% ^{\rm luminosity} \pm 1\% ^{\rm analysis} \pm 0.7\%^{\rm unfolding}$ & \\
 \hline\hline
 B &  $ \pm 1\%$
 & $\pm 1 \%^{t-\rm scale} \pm 0.7\% ^{\rm unfolding}$ & $\mathbf{(20.1 \pm 0.2^{\rm stat} \pm 0.3^{\rm syst})\,GeV^{-2}}$\\
   \hline
 $\frac{{\rm d}\sigma}{{\rm d}t}|_{t=0}$ & $\pm 0.3 \%$
 & $\pm 0.3 \% ^{\rm optics} \pm 4 \% ^{\rm luminosity} \pm 1 \% ^{\rm analysis}$
  & $\mathbf{(503.7 \pm 1.5^{\rm stat} \pm 26.7^{\rm syst})\,mb/GeV^{2}}$ \\
   \hline
   $\int \frac{{\rm d}\sigma}{{\rm d}t}\,{\rm d}t$  & $\pm 0.8 \% ^{\rm extrapolation}$
   & $ \pm 4 \% ^{\rm luminosity} \pm 1\% ^{\rm analysis}$  & \\
   \hline\hline
   $\sigma_{\rm tot}$  & $\pm 0.2\%$
   & $\left(^{+0.8\%}_{-0.2\%}\right)^{\rm (\rho)} \pm 2.7 \%$  & $\mathbf{(98.3 \pm 0.2^{\rm stat} \pm 2.8^{\rm syst})\,mb}$\\
   \hline
   $\sigma_{\rm el} = \int \frac{{\rm d}\sigma}{{\rm d}t}\,{\rm d}t$ & $\pm 0.8 \%$  & $ \pm 5 \%  $  & $\mathbf{(24.8 \pm 0.2^{\rm stat} \pm 1.2^{\rm syst})\,mb}$\\
   \hline
   $\sigma_{\rm inel}$ & $\pm 0.8 \%$  & $\left(^{+2.4\%}_{-1.8\%}\right)$ & $\mathbf{(73.5 \pm 0.6^{\rm stat} ~^{+1.8}_{-1.3} ~^{\rm syst})}$\,{\bf mb} \\
     $\sigma_{\rm inel}$  (CMS)&  & & $(68.0 \pm 2.0^{\rm syst} \pm 2.4^{\rm lumi} \pm 4^{\rm extrap})$\,mb \\
     $\sigma_{\rm inel}$  (ATLAS) &  & & $(69.4 \pm 2.4^{\rm exp} \pm 6.9^{\rm extrap})$\,mb \\
     $\sigma_{\rm inel}$  (ALICE) &  & & $(72.7 \pm 1.1^{\rm model} \pm 5.1^{\rm lumi})$\,mb \\
   \hline
    \multicolumn{2}{l} {$^{\rm (*)}$corrected after unfolding} &
 \multicolumn{2}{r} {$^{\rm analysis}$(includes tagging, acceptance, efficiency, background)}\\
\end{tabular}
 }
\label{tab_results}
\end{center}
\end{table*}


\section{Outlook}
TOTEM foresees taking data in dedicated runs with $\beta^{*} = 90\,$m still in
2011. It is expected that the RP detectors can approach the beam centre as
close as 5 times the transverse beam width. The lowest accessible $|t|$-values
will then be around $0.005\,\rm GeV^{2}$, improving the measurement of the
slope $B$ and the extrapolation of the differential cross-section to $|t| = 0$.
Furthermore, with a running time of at least 5 hours the statistics can be
considerably improved.
Inclusion of the TOTEM inelastic telescopes, T1 and T2,  will
allow a luminosity-independent measurement of
the total cross-section as well as a detailed
study of low-mass diffraction.
If larger values of $\beta^{*}$ (around 1\,km) can be
reached during the year 2012, the acceptance for low-t elastic scattering can
be extended into the Coulomb-Nuclear-Interference region below
$10^{-3}\,\rm GeV^{2}$. A measurement of the $\rho$ parameter might then come
into reach.

\begin{center}
$\ast\ast\ast$
\end{center}
We are indebted to the beam optics development team (A.~Verdier in the initial phase, H.~Burkhardt, G.~M\"{u}ller, S.~Redaelli, J.~Wenninger, S.~White) for
the design, the thorough preparations and the successful commissioning of
the $\beta^{*} = 90\,$m optics.
We congratulate the CERN accelerator groups for the very smooth operation during the optics test run in June 2011. We thank M.~Ferro-Luzzi and the LHC machine coordinators for scheduling the dedicated fills.

We are grateful to CMS for the fruitful and effective collaboration and for providing their luminosity measurements.

We thank M.~Borratynski, E.~Crociani, T.~J\"a\"askel\"ainen, H.~Juntunen, P.~Kaczmarczyk, S.~Podgorski, M.~Polnik, J.~Smajek , J.~Szymanek, T.~Tajakka, P.~Wyszkowski for their help in software development.

This work was supported by the institutions listed on the front page and partially also by NSF (US), the Magnus Ehrnrooth foundation (Finland), the Waldemar von Frenckell foundation (Finland), the Academy of Finland, the OTKA grant NK 73143 (Hungary) and by the NKTH-OTKA grant 74458 (Hungary).


\bibliographystyle{elsarticle-num}

\bibliography{sigma-bibliography-MD}

\begin{thebibliography}{10}
\expandafter\ifx\csname url\endcsname\relax
  \def\url#1{\texttt{#1}}\fi
\expandafter\ifx\csname urlprefix\endcsname\relax\def\urlprefix{URL }\fi
\expandafter\ifx\csname href\endcsname\relax
  \def\href#1#2{#2} \def\path#1{#1}\fi

\bibitem{Amaldi:1973yv}
U.~Amaldi, et~al., {The Energy dependence of the proton proton total
  cross-section for center-of-mass energies between 23 and 53 GeV}, Phys.Lett.
  B44 (1973) 112--118.
\newblock \href {http://dx.doi.org/10.1016/0370-2693(73)90315-8}
  {\path{doi:10.1016/0370-2693(73)90315-8}}.

\bibitem{Amendolia:1973yw}
S.~Amendolia, et~al., {Measurement of the total proton proton cross-section at
  the ISR}, Phys.Lett. B44 (1973) 119--124.
\newblock \href {http://dx.doi.org/10.1016/0370-2693(73)90316-X}
  {\path{doi:10.1016/0370-2693(73)90316-X}}.

\bibitem{Baksay:1978sg}
L.~Baksay, et~al., {Measurement of the Proton Proton Total Cross-Section and
  Small Angle Elastic Scattering at ISR Energies}, Nucl.Phys. B141 (1978)
  1--28.
\newblock \href {http://dx.doi.org/10.1016/0550-3213(78)90331-0}
  {\path{doi:10.1016/0550-3213(78)90331-0}}.

\bibitem{Amaldi:1978vc}
U.~Amaldi, et~al., {Precision measurement of proton proton total cross-section
  at the CERN Intersecting Storage Rings}, Nucl.Phys. B145 (1978) 367.
\newblock \href {http://dx.doi.org/10.1016/0550-3213(78)90090-1}
  {\path{doi:10.1016/0550-3213(78)90090-1}}.

\bibitem{Akimov:1969zzb}
V.~Akimov, et~al., {Measurements of the inelastic proton-proton and
  proton-carbon cross-sections at energies $10^{10}$ to $10^{12}$ eV on board
  the satellites proton 1, 2 and 3}, in: 11th International Cosmic Ray
  Conference (ICRC 1969) 25 Aug - Sep 4, 1969, Budapest, Hungary, 1969, pp.
  211--214.

\bibitem{Yodh:1972fv}
G.~Yodh, Y.~Pal, J.~Trefil, {Evidence for rapidly rising p-p total
  cross-section from cosmic ray data}, Phys.Rev.Lett. 28 (1972) 1005--1008.
\newblock \href {http://dx.doi.org/10.1103/PhysRevLett.28.1005}
  {\path{doi:10.1103/PhysRevLett.28.1005}}.

\bibitem{Dawson:1987iv}
B.~Dawson, {Measurement of the total proton proton cross-section using the
  Fly's Eye}, in: New York 1987, Proceedings, Elastic and diffractive
  scattering, 1987, pp. 337--346.

\bibitem{Battiston:1982su}
R.~Battiston, et~al., {Measurement of the proton\,-\,anti-proton elastic\,and
  total\,cross-section at a center-of-mass\,energy\,of 540\,GeV}, Phys.Lett.
  B117 (1982) 126.
\newblock \href {http://dx.doi.org/10.1016/0370-2693(82)90888-7}
  {\path{doi:10.1016/0370-2693(82)90888-7}}.

\bibitem{Arnison:1983mm}
G.~Arnison, et~al., {Elastic and total cross-section measurement at the CERN
  proton-anti-proton Collider}, Phys.Lett. B128 (1983) 336.
\newblock \href {http://dx.doi.org/10.1016/0370-2693(83)90271-X}
  {\path{doi:10.1016/0370-2693(83)90271-X}}.

\bibitem{Amos:1989at}
N.~Amos, et~al., {Measurement of the $\bar{p} p$ Total Cross-Section at
  $\sqrt{s}$ = 1.8 TeV}, Phys.Rev.Lett. 63 (1989) 2784.
\newblock \href {http://dx.doi.org/10.1103/PhysRevLett.63.2784}
  {\path{doi:10.1103/PhysRevLett.63.2784}}.

\bibitem{Abe:1993xy}
F.~Abe, et~al., {Measurement of the $\bar{p}p$ total cross-section at $\sqrt{s}
  = 546$\,GeV and 1800\,GeV}, Phys.Rev. D50 (1994) 5550--5561.
\newblock \href {http://dx.doi.org/10.1103/PhysRevD.50.5550}
  {\path{doi:10.1103/PhysRevD.50.5550}}.

\bibitem{Sadr:1993jq}
S.~Sadr, et~al., {anti-p p collisions at $s^{1/2}$ = 1.8 TeV: rho, sigma(t),
  and B}, in: International Conference On Elastic And Diffractive Scattering
  (5th Blois Workshop), Providence, Rhode Island, 1993, pp. 59--63.

\bibitem{Avila:1998ej}
C.~Avila, et~al., {A measurement of the proton-antiproton total cross-section
  at $\sqrt{s} = 1.8$\,TeV}, Phys.Lett. B445 (1999) 419--422.
\newblock \href {http://dx.doi.org/10.1016/S0370-2693(98)01421-X}
  {\path{doi:10.1016/S0370-2693(98)01421-X}}.

\bibitem{int:lumi1}
CMS-Collaboration, {Measurement of CMS Luminosity}, Performance Analysis Note
  CMS-PAS-EWK-10-004.

\bibitem{int:lumi2}
CMS-Collaboration, {Absolute luminosity normalization}, Detector Performance
  Note CMS-DP-2011-002 C.

\bibitem{Anelli:2008zza}
G.~Anelli, et~al., {The TOTEM Experiment at the CERN Large Hadron Collider},
  JINST 3 (2008) S08007.
\newblock \href {http://dx.doi.org/10.1088/1748-0221/3/08/S08007}
  {\path{doi:10.1088/1748-0221/3/08/S08007}}.

\bibitem{IPAC:2011}
M.~Deile, et~al., {The First 1 1/2 Years of TOTEM Roman Pot Operation at LHC},
  in: Proceedings of IPAC'11, San Sebastian, Spain, 2011, p. MOPO011.

\bibitem{Verdier:2005av}
A.~Verdier, {TOTEM Optics for LHC V6.5}, CERN-LHC-Project-Note-369-2005.

\bibitem{EggertBlois:2005}
K.~Eggert, {TOTEM Physics}, in: Proceedings of the XIth International
  Conference on Elastic and Diffractive Scattering Towards the High Energy
  Frontiers, Blois, France, May 15-20, 2005.
\newblock \href {http://arxiv.org/abs/hep-ex/0602025}
  {\path{arXiv:hep-ex/0602025}}.

\bibitem{Optics:90memo}
TOTEM-Collaboration, {Early TOTEM running with the 90m Optics},
  CERN-LHCC-2007-013/G-130.

\bibitem{Burkhardt:2010sw}
H.~Burkhardt, S.~White, {High-$\beta^*$ Optics for the LHC},
  CERN-LHC-Project-Note-431-2010.

\bibitem{EPLelastic:2011}
G.~Antchev, et~al., {Proton-proton elastic scattering at the LHC energy of
  $\sqrt{s}=7\,$TeV}, EPL 95 (2011) 41001.
\newblock \href {http://dx.doi.org/10.1209/0295-5075/95/41001}
  {\path{doi:10.1209/0295-5075/95/41001}}.

\bibitem{cmslumi:2011}
CMS-Collaboration, {Absolute Calibration of the CMS Luminosity Measurement:
  Summer 2011 Update}, Performance Analysis Note CMS-PAS-EWK-11-001.

\bibitem{Guryn:2004}
S.~B\protect{\"{u}}ltmann, et~al., {First Measurement of Proton-Proton Elastic
  Scattering at RHIC}, Phys.Lett. B579 (2004) 245--250.

\bibitem{Abe:1993xx}
F.~Abe, et~al., {Measurement of small angle $\bar{p}p$ elastic scattering at
  $\sqrt{s} = 546$ GeV and 1800 GeV}, Phys.Rev. D50 (1994) 5518--5534.
\newblock \href {http://dx.doi.org/10.1103/PhysRevD.50.5518}
  {\path{doi:10.1103/PhysRevD.50.5518}}.

\bibitem{Amos:1988ng}
N.~A. Amos, et~al., {Measurement of $b$, the Nuclear Slope Parameter of the
  $p\bar{p}$ Elastic Scattering Distribution at $\sqrt{s}$ = 1800-GeV},
  Phys.Rev.Lett. 61 (1988) 525.
\newblock \href {http://dx.doi.org/10.1103/PhysRevLett.61.525}
  {\path{doi:10.1103/PhysRevLett.61.525}}.

\bibitem{Amos:1990fw}
N.~A. Amos, et~al., {$\bar{p}p$ Elastic Scattering at $\sqrt{s} = 1.8 TeV$ from
  $|t| = 0.034 GeV/c^{2}$ to $0.65 GeV/c^{2}$}, Phys.Lett. B247 (1990)
  127--130.
\newblock \href {http://dx.doi.org/10.1016/0370-2693(90)91060-O}
  {\path{doi:10.1016/0370-2693(90)91060-O}}.

\bibitem{Amos:1992zn}
N.~A. Amos, et~al., {$\bar{p}p$ Elastic Scattering At $\sqrt{s}$ = 1020 GeV},
  Nuovo Cim. A106 (1993) 123--132.
\newblock \href {http://dx.doi.org/10.1007/BF02771512}
  {\path{doi:10.1007/BF02771512}}.

\bibitem{Brandt:2010zz}
A.~Brandt, {D0 Measurement of the Elastic $ p \bar p$ Differential Cross
  Section for $0.25 < |t| < 1.2\,\rm GeV^2$ at $s^{(1/2)} = 1.96\,\rm TeV$},
  PoS DIS2010 (2010) 059, ~FERMILAB-CONF-10-547.

\bibitem{Cudell:2002}
J.~Cudell, et~al., {Benchmarks for the Forward Observables at RHIC, the
  Tevatron-Run II, and the LHC}, Phys.Rev.Lett. 89 (2002) 201801.

\bibitem{CMSint:2011}
CMS-Collaboration, {Inelastic pp cross section at 7 TeV}, Performance Analysis
  Note CMS-PAS-FWD-11-001.

\bibitem{Aad:2011eu}
G.~Aad, et~al., {Measurement of the Inelastic Proton-Proton Cross-Section at
  $\sqrt{s}=7$\,TeV with the ATLAS Detector}, arXiv hep-ex/1104.0326.

\bibitem{ALICE:2011}
M.~Poghosyan, {Diffraction dissociation in proton-proton collisions at
  $\sqrt{s} = 0.9\,\rm TeV$, 2.76\,TeV and 7\,TeV with ALICE at the LHC}, arXiv
  hep-ex/1109.4510.

\bibitem{PDG:2010}
K.~Nakamura, et~al. (Particle Data~Group), J. Phys. G 37 (2010) 075021.

\bibitem{Block:2011nr}
M.~Block, {Ultra-high Energy Predictions of Proton-Air Cross Sections from
  Accelerator Data: an Update}, arXiv hep-ph/1109.2940.

\end{thebibliography}

\end{document}